\documentclass[twocolumn]{aastex631}
\usepackage{hyperref, natbib, color, subfigure, amsmath, soul, verbatim}
\usepackage[T1]{fontenc}
\graphicspath{{./}{figures/}}

\newcommand{\hbeta}{H{$\beta$}\,}
\newcommand{\halpha}{H{$\alpha$}\,}
\newcommand{\hgamma}{H{$\gamma$}\,}
\newcommand{\lya}{Ly\,$\alpha$\,}
\newcommand{\CIII}{\ion{C}{3}]\,}

\newcommand{\CaII}{\ion{Ca}{2}\,}
\newcommand{\OII}{[\ion{O}{2}]\,3727}
\newcommand{\NeV}{[\ion{Ne}{5}]\,}
\def\CIV{\ion{C}{4}\,}
\def\MgII{\ion{Mg}{2}\,}
\def\FeII{\ion{Fe}{2}\,}
\def\OIII{[\ion{O}{3}]\,5007\,}
\def\HeII{\ion{He}{2}\,1640\,}
\def\SiIV{\ion{Si}{4}\,}
\def\NII{[\ion{N}{2}]\,6585\,}

\def\kms{\textit{$\rm km\,s^{-1}$}}

\def\eV{\textit{$\rm eV$}}

\def\Mbh{\textit{$M_{\rm BH}$}}

\def\LLEdd{\textit{$L/L_{\rm Edd}$}}

\def\zsys{\textit{$z_{\rm sys}$}}
\def\agnvac{Juneau et al., in prep}
\shorttitle{A Catalog of DESI DR1 Quasars}
\shortauthors{Wu \& Shen}

\begin{document}

\title{A Catalog of Spectral Properties for DESI DR1 Quasars}

\author[0000-0003-4202-1232]{Qiaoya Wu}
\email{qiaoyaw2@illinois.edu}
\affiliation{Department of Astronomy, University of Illinois at Urbana-Champaign, Urbana, IL 61801, USA}

\author[0000-0003-1659-7035]{Yue Shen}
\email{shenyue@illinois.edu}
\affiliation{Department of Astronomy, University of Illinois at Urbana-Champaign, Urbana, IL 61801, USA}
\affiliation{National Center for Supercomputing Applications, University of Illinois at Urbana-Champaign, Urbana, IL 61801, USA}

\begin{abstract}
   The Dark Energy Spectroscopic Instrument (DESI) is carrying out a five-year spectroscopic survey designed to map the large-scale structure of the Universe and obtain spectra for millions of galaxies and quasars over a wide redshift range. We present a value-added catalog of spectral properties for 1,717,774 unique quasars (1,774,723 spectra) in the DESI Data Release 1. We perform a uniform spectral decomposition of the quasar continuum, host galaxy emission (at $z<1$), and \FeII\ emission, and measure the line properties of major broad and narrow emission lines from Ly$\alpha$ to H$\alpha$ covered by DESI spectra. We also provide single-epoch black hole mass estimates (from H$\beta$, \MgII, and \CIV), bolometric luminosities, and Eddington ratios for these quasars. We provide systemic redshifts that account for the intrinsic velocity shifts of individual lines; they agree with the DESI pipeline redshifts to within 500~km s$^{-1}$ for 88.6\% of the sample, and are most useful at $z>1$ where the pipeline redshift is heavily constrained by blueshifted high-ionization lines. We also provide local fits to several optical coronal lines, including \NeV~$\lambda3426$, [\ion{Fe}{7}]~$\lambda5722$, [\ion{Fe}{7}]~$\lambda6088$, and [\ion{Fe}{11}]~$\lambda7893$. We show that several emission-line ratios, including those involving coronal lines, vary systematically with Eddington ratio. The catalog is intended as a community resource for studies of quasar demographics, black hole accretion, emission-line physics, and the connection between the broad-, coronal-, and narrow-line regions.
\end{abstract}


\section{Introduction}\label{sec:intro}

Large, homogeneous spectroscopic surveys have transformed quasar astrophysics by enabling detailed demographic studies and systematic measurements of emission‐line and continuum properties across cosmic time. The Sloan Digital Sky Survey \citep[SDSS;][]{York2000} established this paradigm, producing successively larger quasar catalogs through BOSS and eBOSS and culminating in DR16Q \citep{Lyke_etal_2020}. Value-added measurements from SDSS spectra, such as emission-line properties, quasar luminosities, and single-epoch (SE) black hole (BH) masses, as well as improved systemic redshifts, have become foundational community resources \citep[e.g.,][]{Shen_etal_2011, Rakshit_etal_2020, Wu&Shen2022}.

The Dark Energy Spectroscopic Instrument \citep[DESI;][]{Levi_etal_2013_DESI} survey builds on this legacy and represents a major advance in wide-area spectroscopic surveys. Installed on the Mayall 4-meter Telescope at Kitt Peak National Observatory, DESI is the result of an international collaboration effort to deliver efficient, high-throughput, and uniformly processed spectroscopy for tens of millions of targets. DESI Data Release 1 \citep[DR1;][]{DESI_DR1_2025} comprises the first 13 months of Main Survey observations, together with uniformly reprocessed Survey Validation data, and delivers high-confidence redshifts for $\sim 18.7$ million objects, including $\sim 1.7$ million quasars. As the largest spectroscopic compilation to date, DESI DR1 provides an unprecedented data set for studying quasar spectral diversity, improving systemic redshift measurements, and enabling statistical investigations of SMBH growth, outflows, and the quasar luminosity function.

In this work, we present a value-added catalog of spectral properties for DESI DR1 quasars. This paper is organized as follows. Section~\ref{sec:data} describes the DESI DR1 quasar sample and spectroscopy, and Section~\ref{sec:method} presents the spectral fitting procedure and derived quantities. In Section~\ref{sec:sample_prop}, we validate the catalog using systemic redshifts and comparisons with SDSS, and present how emission-line ratios trend with Eddington ratio. We summarize our results in Section~\ref{sec:conclude}. Throughout this paper, we adopt a flat $\Lambda$CDM cosmology with $\Omega_{\Lambda}=0.7$, $\Omega_{M}=0.3$, and $H_0=70\,{\rm km\,s^{-1}\,Mpc^{-1}}$.

\section{Data} \label{sec:data}

\begin{figure}
\centering
    \includegraphics[width=\linewidth]{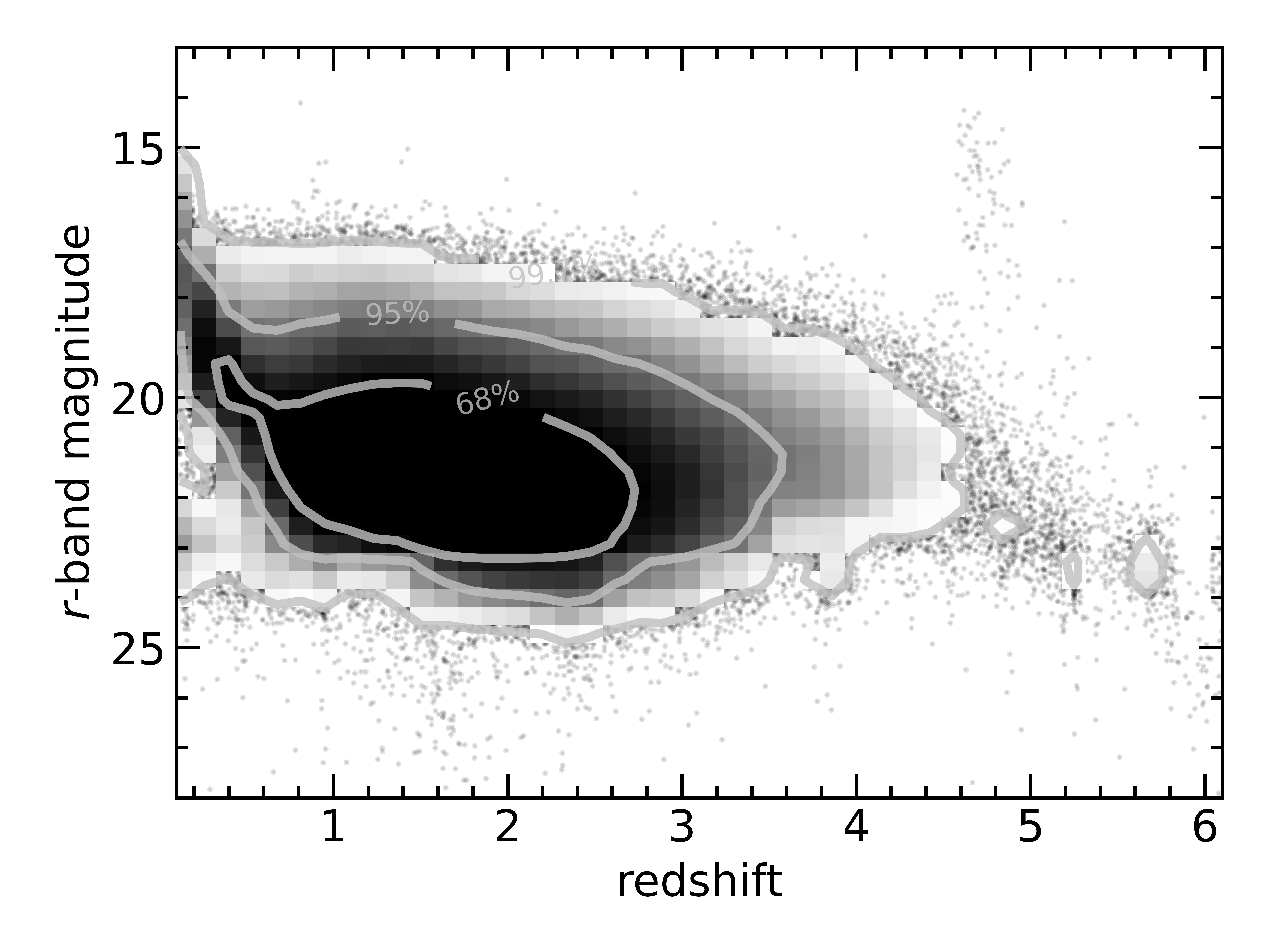}
    \caption{Distribution of DESI DR1 quasars in the magnitude-redshift plane. Apparent $r$-band PSF magnitudes are computed as $m_r = 22.5-2.5\log_{10}({\rm flux}_r)$ from the DESI DR1 targeting photometry catalog (\url{https://data.desi.lbl.gov/doc/releases/dr1/vac/lsdr9-photometry/}). The contours indicate the $1\sigma$, $2\sigma$, and $3\sigma$ density levels.}\label{fig:magr_z}
\end{figure}

\begin{figure}
\centering
    \includegraphics[width=\linewidth]{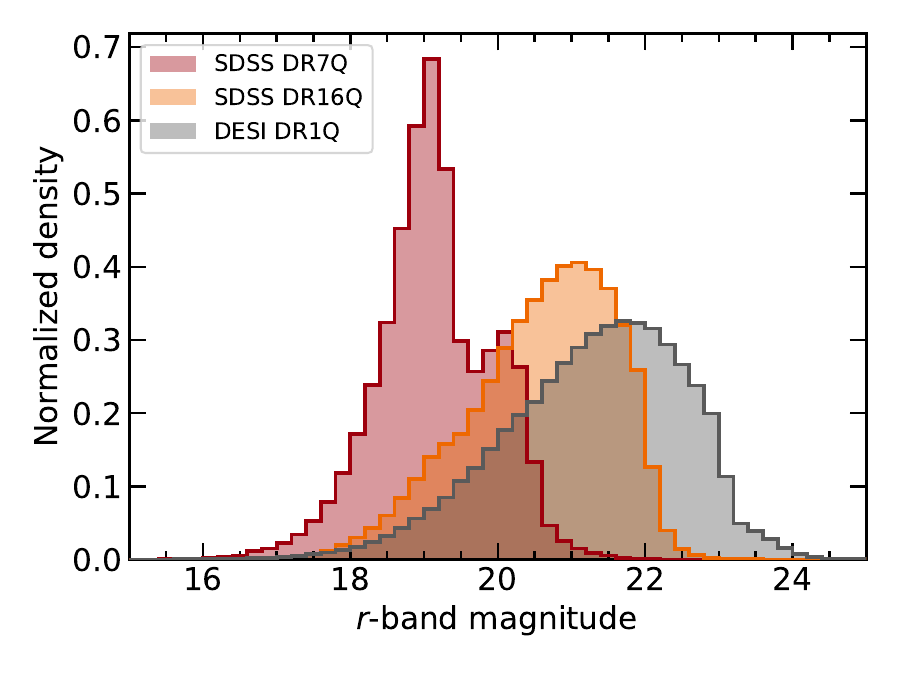}
    \caption{Normalized $r$-band magnitude distributions for SDSS DR7Q, SDSS DR16Q, and DESI DR1Q.}
    \label{fig:magr_compare}
\end{figure}

Quasar targets in DESI are selected primarily from optical imaging in $g$, $r$, $z$ bands, and WISE W1, W2 using a random forest algorithm to improve quasar selection efficiency \citep{Chaussidon_etal_2023_DESI, Alexander_etal_2023}. 
The main-survey selection achieves high quasar purity at magnitude range $16.5<r<23$ with more than $\sim 200$ quasars per deg$^2$ (with $60$ deg$^{-2}$ at $z>2.1$) \citep{Chaussidon_etal_2023_DESI}. Detailed algorithms for the DESI target selections and spectral reduction are presented in \citet{Myers+2023, Chaussidon_etal_2023_DESI, Guy_etal_2023}.
DESI deploys 5,000 fiber-positioning robots with ten corresponding spectrographs with three arms: blue (B; $3600-5800$ \AA), red (R; $5760-7620$ \AA), and near-infrared (Z; $7520-9824$ \AA). We download the HEALPix-coadded one-dimensional spectra produced by the DESI collaboration with sky subtraction and flux calibration from the SPARCL database \footnote{https://datalab.noirlab.edu/data/desi}. The combined wavelength coverage is $3600-9824$ \AA\ with resolution $R\sim2000-5200$ (arm-dependent) and a fixed pixel size of $0.8$ \AA\ \citep{Guy_etal_2023, DESI_DR1_2025}.

We begin with the DESI DR1 AGN/QSO Value-Added Catalog \footnote{https://data.desi.lbl.gov/doc/releases/dr1/vac/agnqso/}, which includes all DR1 spectra classified as extragalactic (\agnvac). 
We use {\tt SPECTYPE=QSO} to select out targets that the Redrock pipeline (Bailey et al., in prep.) assigned the classification, and use {\tt AGN\_MASKBITS} for objects selected by the machine-learning algorithm QuasarNet and the \MgII\ doublet post-processing pipeline \citep{Chaussidon_etal_2023_DESI, DESI_EDR, DESI_DR1_2025}. This yields a total of 1,717,774 unique quasars with 1,774,723 spectra. Figure~\ref{fig:magr_z} shows the distribution of the DESI DR1 quasars in the magnitude--redshift plane. Figure~\ref{fig:magr_compare} compares the $r$-band magnitude distributions of SDSS DR7Q \citep{Schneider_etal_2010, Shen_etal_2011}, SDSS DR16Q \citep{Lyke_etal_2020, Wu&Shen2022}, and DESI DR1Q, illustrating the progressively fainter quasar populations reached by successive large spectroscopic surveys over the past two decades. The SDSS DR7Q distribution peaks near $m_r\sim19$ with a relatively sharp faint-end cutoff toward fainter magnitudes due to the relatively bright, color-selected flux limit of the SDSS-I/II quasar survey \citep{Richards_2002_SDSS_target}. The SDSS DR16Q extends roughly two magnitudes fainter, consistent with the deeper eBOSS quasar target selection \citep{Myers_2015_SDSS_target}. DESI DR1Q reaches the faintest magnitudes of the three, with a distribution extending to $m_r\sim24$, which reflects increasingly sophisticated selection algorithms, larger telescope aperture and/or better spectrograph sensitivity, and the consequence of including high-$z$ \lya-forest quasar targets \citep{Chaussidon_etal_2023_DESI, Myers+2023}. 

\section{Methodology} \label{sec:method}

\begin{figure*}
\centering
    \includegraphics[width=\linewidth]{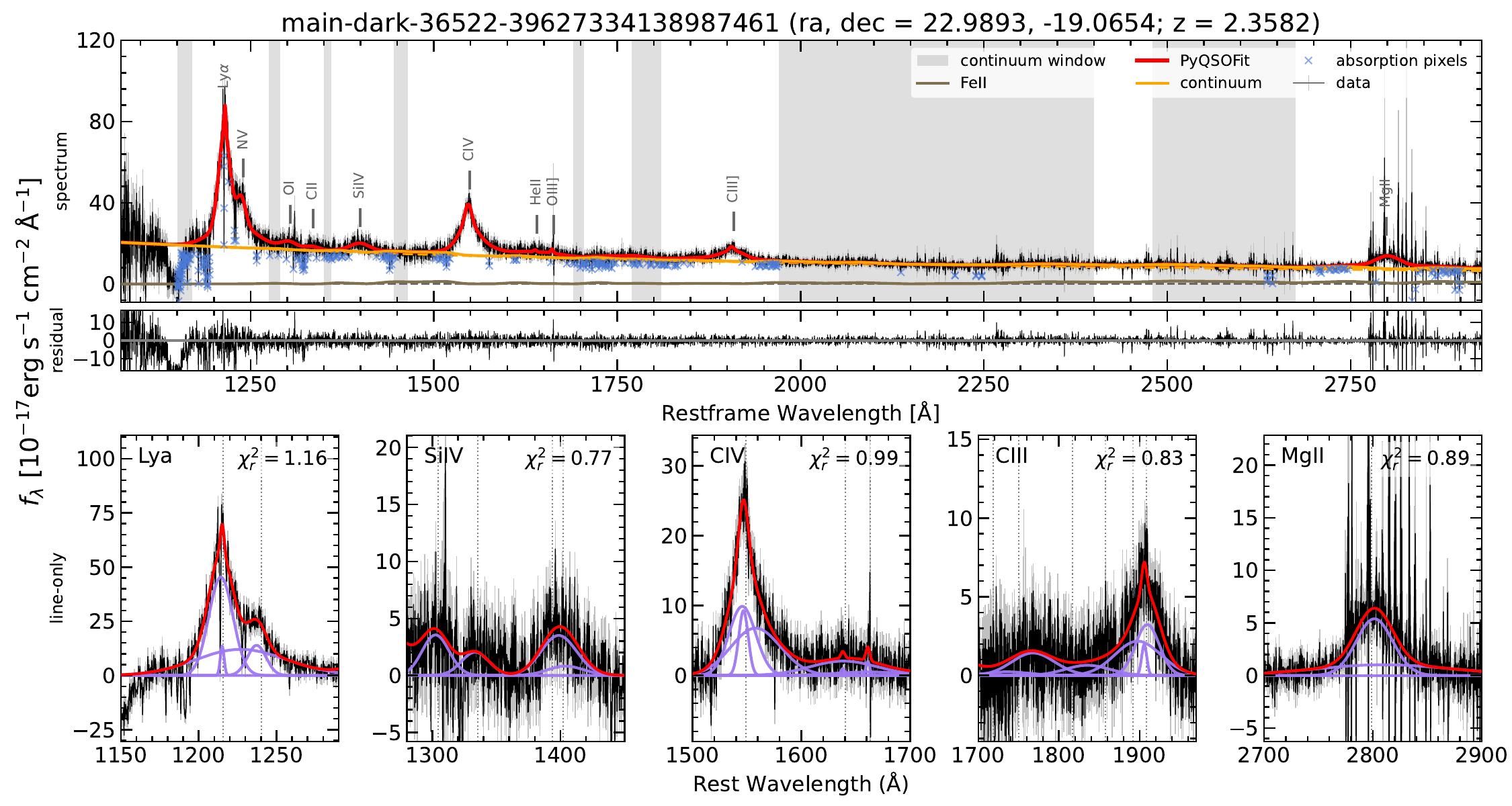}
    \caption{
    Example PyQSOFit result for a DESI DR1 quasar in the rest frame (main-dark-36522-39627334138987461, $z = 2.358$). Top: DESI spectrum (black) with the best-fit total model (red), pseudo-continuum (orange; power law plus polynomial plus \FeII), and UV \FeII\ emission (brown). Gray regions mark the continuum windows used in the continuum fit. Blue crosses mark pixels rejected as absorption. Bottom: continuum-subtracted zoom-ins on the fitted line complexes (\lya+\ion{N}{5}, \SiIV, \CIV, \CIII, \MgII). The red curve is the total line model, and the purple curves are its individual Gaussian components. The reduced $\chi^2$ of each complex fit is given in its panel.}
    \label{fig:QA}
\end{figure*}

\begin{figure}
\centering
    \includegraphics[width=\linewidth]{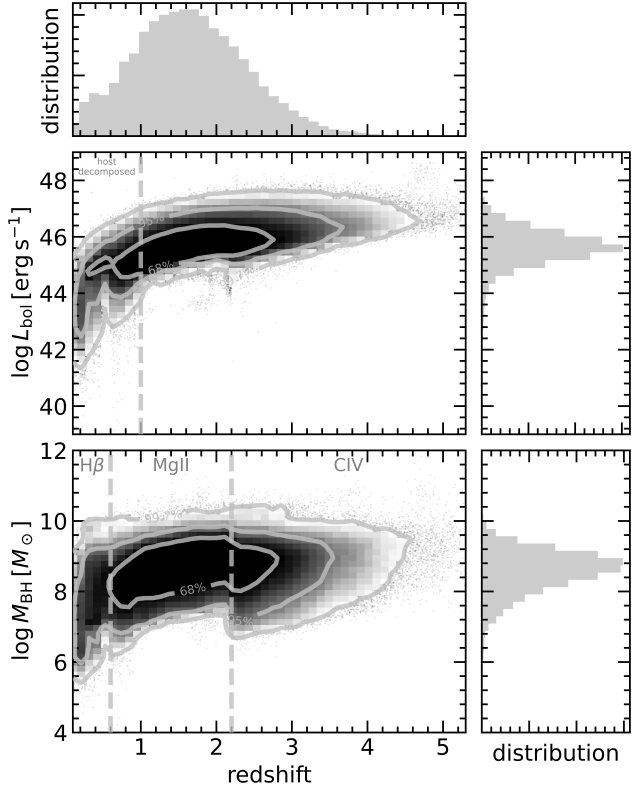}
    \caption{Distributions of DESI DR1 quasar BH mass (\Mbh) and bolometric luminosity ($L_{\rm bol}$) as functions of redshift for objects with {\tt SNR\_ALL\_MEDIAN}$>1$, with the side/top panels showing the corresponding one-dimensional PDFs.
    }
    \label{fig:mbh_lbol_z}
\end{figure}

We measure spectral properties for these quasars following the practice in earlier work \citep{Shen_etal_2011, Shen_etal_2019b, Wu&Shen2022, Wu&Shen2023_DESI_EDR} using the publicly available code PyQSOFit \citep{PyQSOFit}. Each quasar spectrum was first corrected for the Milky Way extinction using the dust maps \citep{SFD, Schlafly&Finkbeiner2011} and then shifted to the rest frame for spectral analysis. For quasars at redshift $z<1.0$, we perform the host-galaxy decomposition with priors to remove stellar contamination \citep{Ren_etal_2024}. Then the continuum is modeled with a combination of a power law, a third-order polynomial, and \FeII\ templates, fitted to line-free regions of the spectrum. Emission and absorption lines are then modeled with multiple Gaussian components, allowing for measurements of peak wavelength, line fluxes, luminosities, full width at half maximum (FWHM), equivalent widths (EW), and centroid wavelength. More details on the spectral fitting can be found in \citet{Wu&Shen2022}. We adopt a Monte Carlo approach with 10 randomly perturbed spectra to estimate the uncertainties. Figure~\ref{fig:QA} shows an example of our spectral fits.

Bolometric luminosities are estimated from continuum luminosity at 1350~\AA, 3000~\AA, and 5100~\AA using empirical bolometric corrections $\rm BC =$ 3.81, 5.15, and 9.26, respectively \citep{Richards_etal_2006,Shen_etal_2011}. 
BH masses were estimated from the continuum and broad emission lines using SE virial relations \citep{Vestergaard&Peterson2006}. We adopt the same fiducial BH mass recipes on \hbeta, \MgII\ and \CIV\ as in previous work \citep{Shen_etal_2011}: 
\begin{equation}
    \log\left(\frac{M_{\rm BH}}{M_\odot}\right)=a + b\log\left(\frac{\lambda L_\lambda}{10^{44}\, {\rm erg\,s^{-1}}}\right)+2\log\left(\frac{{\rm FWHM}}{{\rm km\,s^{-1}}}\right)\ ,
\end{equation}
where $(a,b)=(0.91, 0.50)$ for \hbeta, $(0.74, 0.62)$ for \MgII\ and $(0.66, 0.53)$ for \CIV. 
For \CIV\ we use the FWHM of the broad profile after rejecting Gaussian components that contribute less than 5\% of the total broad-line flux. 
We include the BH mass estimation based on all three lines (\hbeta, \MgII, and \CIV) and the fiducial BH mass based on the redshift (we adopt the \MgII\ mass for $0.6\le z_{\rm pip}\le2.2$ when available, otherwise the \hbeta\ mass for $z_{\rm pip}\leq0.8$ or the \CIV\ mass for $z_{\rm pip}\geq1.8$). The fiducial $L_{\rm bol}$ follows the same redshift rule, using $\nu L_{1350}$, $\nu L_{3000}$, $\nu L_{5100}$ respectively. These fiducial BH masses and bolometric luminosities are used to calculate the Eddington ratio with 
$L_{\rm bol}/L_{\rm Edd}\equiv L_{\rm bol}/(1.26\times10^{38}\,\Mbh)$. The distributions of the bolometric luminosity and mass are demonstrated in Figure~\ref{fig:mbh_lbol_z}. A small discontinuity in the virial BH masses is visible near $z\sim2$, where the fiducial mass estimator transitions from \MgII\ to \CIV.  
This offset is substantially reduced when the sample is restricted to higher-S/N spectra, suggesting that it may arise from increased emission-line measurement uncertainties for the fainter DESI quasars, potentially combined with a mass-dependent systematic difference between the \MgII- and \CIV-based virial estimators at low BH masses.

To derive a systemic redshift from the spectral fit, we follow our earlier work \citep{Shen_etal_2016b, Wu&Shen2022, Wu&Shen2023_DESI_EDR}. Each line included in the fit provides an estimate of the systemic redshift, after accounting for the average intrinsic velocity shifts of different lines and the luminosity dependence of the two high-ionization lines \CIV\ and \SiIV. We estimate $z_{\rm sys,line}$ for each line (broad \hbeta, \OIII, \CaII, \OII, \MgII, broad \CIII, \CIV, and \SiIV) with a $>3\sigma$ flux detection and spectral coverage $N_{\rm pix}>N_{\rm pix,max}/2$. For objects with at least three valid lines, we reject $z_{\rm sys, line}$ values deviating by more than 3 median absolute deviations from the median, and then restore \OIII, \CaII\ and \OII\ if they are detected at $>3\sigma$. The adopted $z_{\rm sys}$ is the inverse-variance-weighted mean of the remaining lines.
We then decide between $z_{\rm sys}$ and the DESI pipeline redshift $z_{\rm pip}$ ({\tt Z\_FIT}, which uses the \lya\ BAO redshift where available). We find that $z_{\rm pip}$ is generally more reliable than \zsys\ for low-redshift objects ($z \lesssim 1$) and for spectra with very low signal-to-noise ($SNR \lesssim 1$). In these regimes, host-galaxy features and/or low-SNR template degeneracies can bias line-centroid–based systemic redshift recipes. 
For objects with large discrepancies, we adopt the weighted $z_{\rm sys}$ for objects with a SDSS DR16Q counterpart unless $z_{\rm pip}$ is closer to the DR16Q systemic redshift, and for objects without a counterpart in DR16Q we use $z_{\rm pip}$ if a \lya\ redshift is available and the weighted $z_{\rm sys}$ otherwise. 

The spectral measurements and derived quantities are described in Table \ref{tab:fits_catalog}.

\section{Sample properties} \label{sec:sample_prop}

\subsection{Improved systemic redshifts}\label{sec:zsys}

\begin{figure}
\centering
    \includegraphics[width=\linewidth]{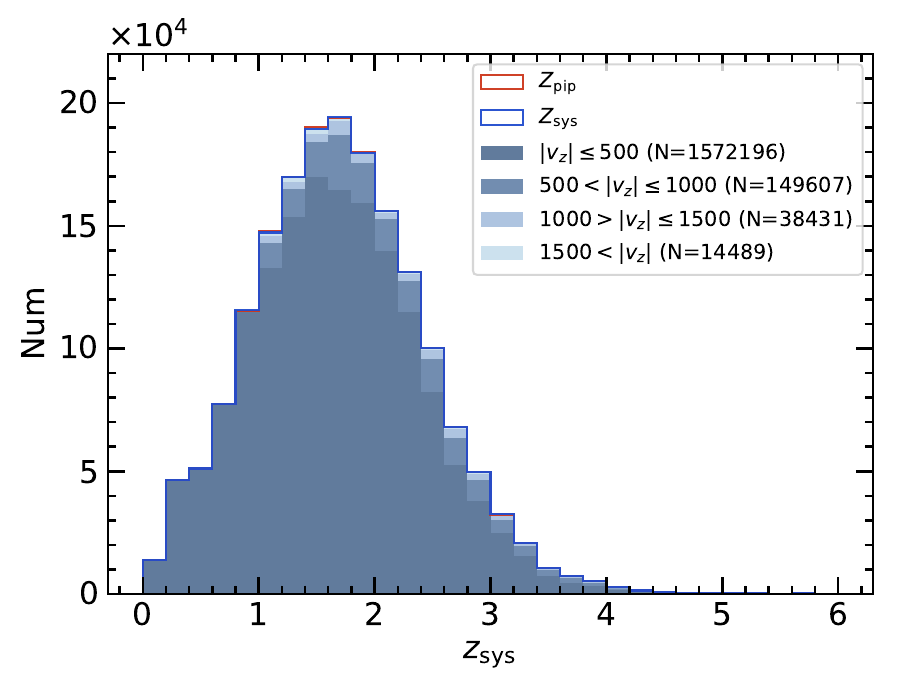}
    \caption{Redshift distributions of DESI DR1 quasars with different velocity discrepancies between the pipeline redshift and our systemic redshift.
    The histogram of \zsys\ and $z_{\rm pip}$ follow the same redshift distribution across the full sample, with $88.6\%$ of quasars having $|v_z|\leq500$ \kms, indicating overall good agreement between the two redshift estimates, while only a small fraction of sources show larger offsets.}
    \label{fig:zsys_zpip}
\end{figure}

\begin{figure*}
\centering
    \includegraphics[width=\linewidth]{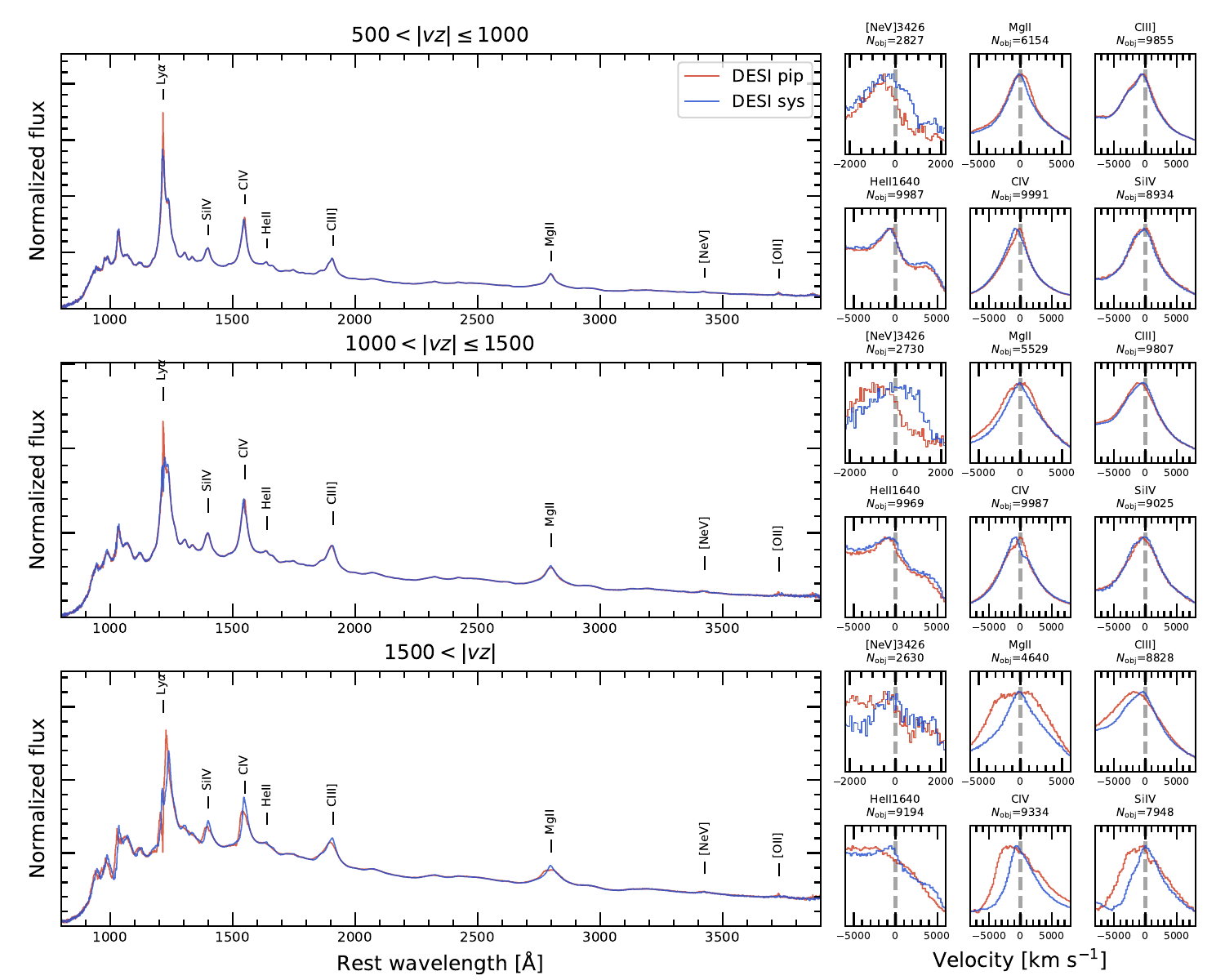}
    \caption{
      Median composite spectra around UV emission lines of DESI DR1 $z>1.5$ quasar subsets (10k objects each) with large discrepancies between the default pipeline redshift and the systemic redshift.
      }\label{fig:coadd_spec}
\end{figure*}

DESI spectra are processed by the Redrock pipeline, which assigns a spectroscopic classification and redshift to each target \citep{Guy_etal_2023, DESI_DR1_2025}. For quasar candidates, DESI further incorporates the QuasarNet classifier/redshift estimator and an \MgII-based recipe to flag cases where the initial Redrock classification/redshift may be ambiguous or affected by template degeneracies to improve the robustness of the adopted quasar redshifts \citep{Chaussidon_etal_2023_DESI, Alexander_etal_2023}. For our spectral fits, we also adopt the updated redshifts from the DESI Year 1 \lya\ BAO pipeline whenever available \citep{Filbert+2024, Martini+2025}.
Overall, the pipeline redshifts perform well for the majority of DESI DR1 quasars. However, redshift inference can be affected by the diversity of quasar spectral properties and by systematic velocity shifts of broad emission lines, especially in regimes where the observed spectrum is dominated by rest-frame UV features. 
As shown in Figure~\ref{fig:coadd_spec}, we compare the composite spectra constructed using the pipeline redshift ($z_{\rm pip}$) and the improved systemic redshift (\zsys) for a random 10k subset. For objects with small velocity discrepancies, the two stacks are nearly identical, indicating that the pipeline redshifts are highly reliable. As the redshift discrepancy increases, systematic differences become more apparent. The \zsys\ composite spectra exhibit narrower and sharper line profiles compared to those constructed using $z_{\rm pip}$. In particular, high-ionization lines such as \CIV\ are known to exhibit luminosity-dependent blueshifts relative to their rest-frame wavelengths \citep{Coatman+2016, Shen_etal_2016b}, which the \zsys-based stacks clearly recover. 
This suggests that the pipeline redshift determination, which relies on the observed line peaks, can be biased by these blueshifted features. In addition, objects with large velocity discrepancies display absorption features and asymmetric line profiles, indicating the presence of broad absorption line quasars in this sample. In such cases, the pipeline is more prone to misidentifying the redshift due to partially absorbed emission profiles.
Our improved systemic redshifts mitigate such biases by anchoring the redshift to both the emission line features and the quasar luminosity, resulting in better alignment of emission-line centroids across the stacked spectra.

We compare \zsys\ to $z_{\rm pip}$ across the full quasar sample, grouped by their velocity difference $v_z$  in Figure~\ref{fig:zsys_zpip}. 
The two redshifts agree well across the full redshift range. The majority of objects exhibit small velocity differences: 1,572,196 quasars (88.6\%) have $|v_z|\leq500$\,\kms, while an additional 149,607 objects (8.4\%) fall within $500<|v_z|\leq1000$\,\kms. Only 38,431 quasars (2.2\%) show offsets in the range $1000<|v_z|\leq1500$\,\kms, and a small fraction of 14,489 objects (0.8\%) exhibit larger discrepancies with |$v_z|>1500$\,\kms. The overall consistency of the two redshift distributions indicates that the DESI pipeline redshifts are generally robust for quasar redshift characterization. The remaining discrepancies likely arise from known challenges in quasar redshift determination, particularly at redshifts where only broad rest-frame UV emission lines are available.

\subsection{Coronal lines}

\begin{table}
\caption{Coronal lines fitted in the DESI DR1 quasar spectra. $\chi$ is the ionization potential required to produce the ion. $N_{\rm fit}$ is the number of spectra covering the line and $N_{\rm det}$ is the number of detections.}
\centering
\begin{tabular}{lrrrrr}
\hline
Line & $\lambda$ (\AA) & $\chi$ (eV) & $N_{\rm fit}$ & $N_{\rm det}$ & $f_{\rm det}$ (\%) \\
\hline
{[\ion{Ne}{5}]}  & 3346.82 &  97.1 & 1,079,952 &  9,660 & 0.89 \\
{[\ion{Ne}{5}]}  & 3426.85 &  97.1 & 1,019,083 & 96,591 & 9.48 \\
{[\ion{Fe}{7}]}  & 3587.34 &  99.1 &   897,370 &  1,606 & 0.18 \\
{[\ion{Fe}{7}]}  & 3759.99 &  99.1 &   778,761 &  3,846 & 0.49 \\
{[\ion{Fe}{14}]} & 5304.34 & 361.0 &   199,516 &    261 & 0.13 \\
{[\ion{Fe}{7}]}  & 5722.29 &  99.1 &   139,877 &  4,240 & 3.03 \\
{[\ion{Fe}{7}]}  & 6088.69 &  99.1 &   107,900 &  7,416 & 6.87 \\
{[\ion{Fe}{10}]} & 6376.26 & 233.6 &    90,548 &    332 & 0.37 \\
{[\ion{S}{12}]}  & 7613.10 & 504.8 &    28,436 &     60 & 0.21 \\
{[\ion{Fe}{11}]} & 7893.97 & 262.1 &    18,459 &    810 & 4.39 \\
\hline
\end{tabular}\label{tab:CL_table}
\end{table}

\begin{figure*}
\centering
    \includegraphics[width=\linewidth]{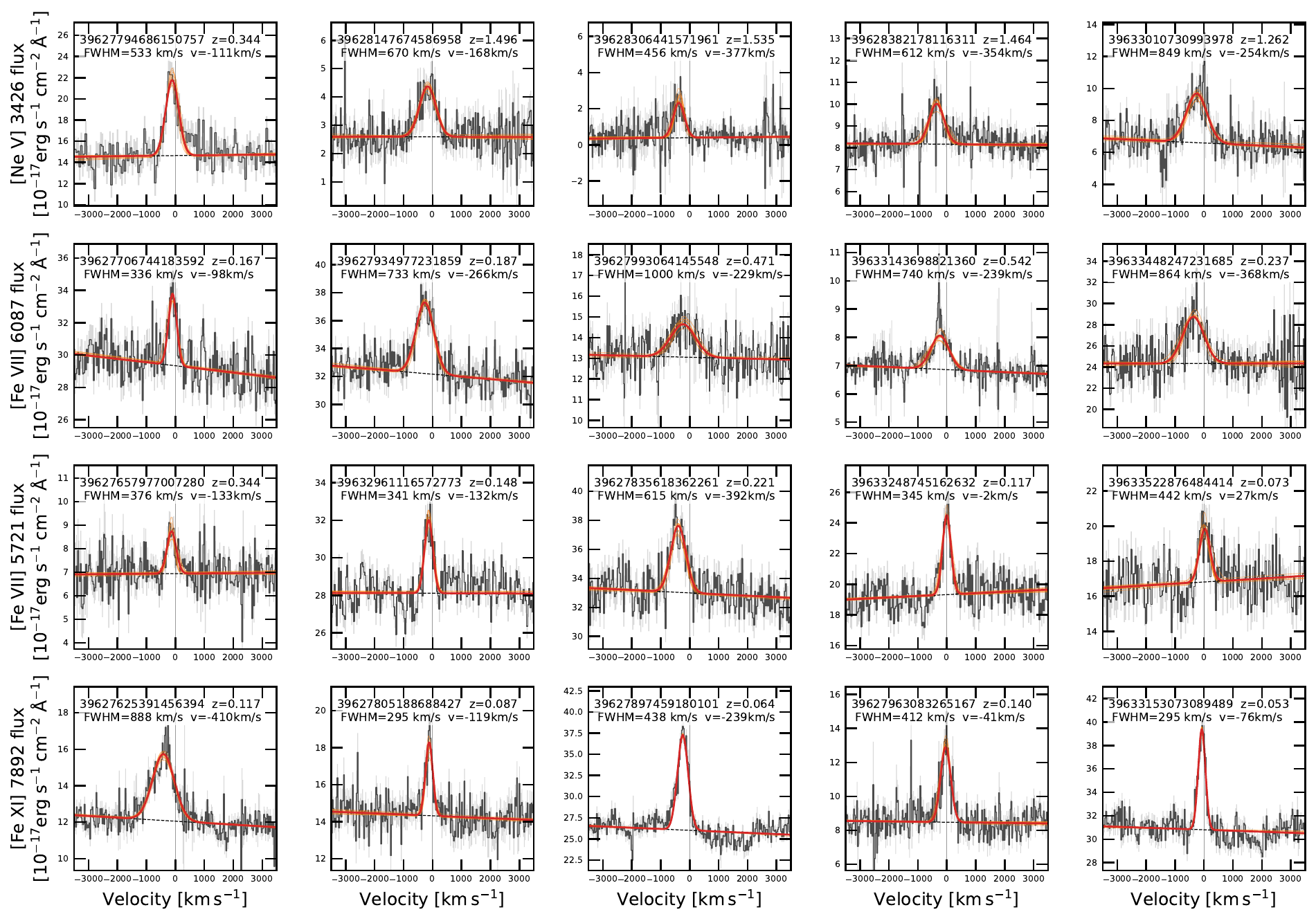}
    \caption{Examples of local fits to coronal emission lines in DESI DR1 quasar spectra. Each row presents five randomly selected examples for a given coronal line: [\ion{Ne}{5}]~$\lambda3426$, [\ion{Ne}{5}]~$\lambda3346$, [\ion{Fe}{7}]~$\lambda6087$, and [\ion{Fe}{7}]~$\lambda5721$ from top to bottom. The observed spectra are shown in black, while the red curves represent the best-fit models consisting of a local continuum and a single Gaussian component. Each panel is annotated with the object information and basic line properties.}\label{fig:CL_demo}
\end{figure*}

Coronal lines (CLs) are forbidden transitions from highly ionized species with ionization potentials of IP$\gtrsim100~\eV$. Producing these ions requires an intense extreme-UV or soft X-ray radiation field, making CLs sensitive probes of the hardest portion of the ionizing spectral energy distribution (SED) \citep{Ferguson+1997, Reunanen+2003}. Such hard ionizing continua are difficult to produce in normal stellar populations, and strong CL emission is therefore most commonly associated with active galactic nucleus (AGN) activity \citep{Reunanen+2003, Goulding&Alexander_2009}. 
High-ionization transitions such as [\ion{Ne}{5}] have been successfully used to identify accreting SMBHs in galaxies for which conventional optical AGN signatures are weak or obscured \citep{Goulding&Alexander_2009, Gilli+2010_NeV, Negus_etal2023, Acuna_2026_NeV}. Alternatively, some studies indicate that fast radiative shocks between high-velocity clouds and the ISM may also contribute to CL production \citep{Allen+2008_MAPPINGS, MullerS_etal2011}.

We search for and fit a set of optical coronal lines (CLs; Table~\ref{tab:CL_table}) in the DESI quasar spectra over the redshift ranges where each transition is covered. Each line is fitted locally within a $\pm5000~\kms$ window centered on its expected wavelength using a linear continuum plus a single Gaussian profile. We classify a line as detected when the fitting window is fully covered, the Gaussian model is preferred over a continuum-only model by $\Delta{\rm BIC}>10$, the line has ${\rm S/N}\ge3$, and none of the fitted parameters reaches its imposed boundary.  Figure~\ref{fig:CL_demo} presents several CL-detected examples. 

Systematic CL measurements provide several opportunities for quasar studies. Their high ionization potentials make them powerful diagnostics of the EUV/soft-X-ray continuum, which is otherwise poorly constrained observationally. Ratios involving [Ne V] and lower-ionization lines can also improve AGN identification at redshifts where standard optical diagnostics are inaccessible. In addition, CL widths and velocity shifts provide complementary tracers of highly ionized outflows. Combined with broad-line properties, BH mass, and Eddington ratio, these measurements can help connect the BLR, CLR, and NLR to the underlying accretion state.

\subsection{Line ratios}

\begin{figure*}
  \includegraphics[width=\linewidth]{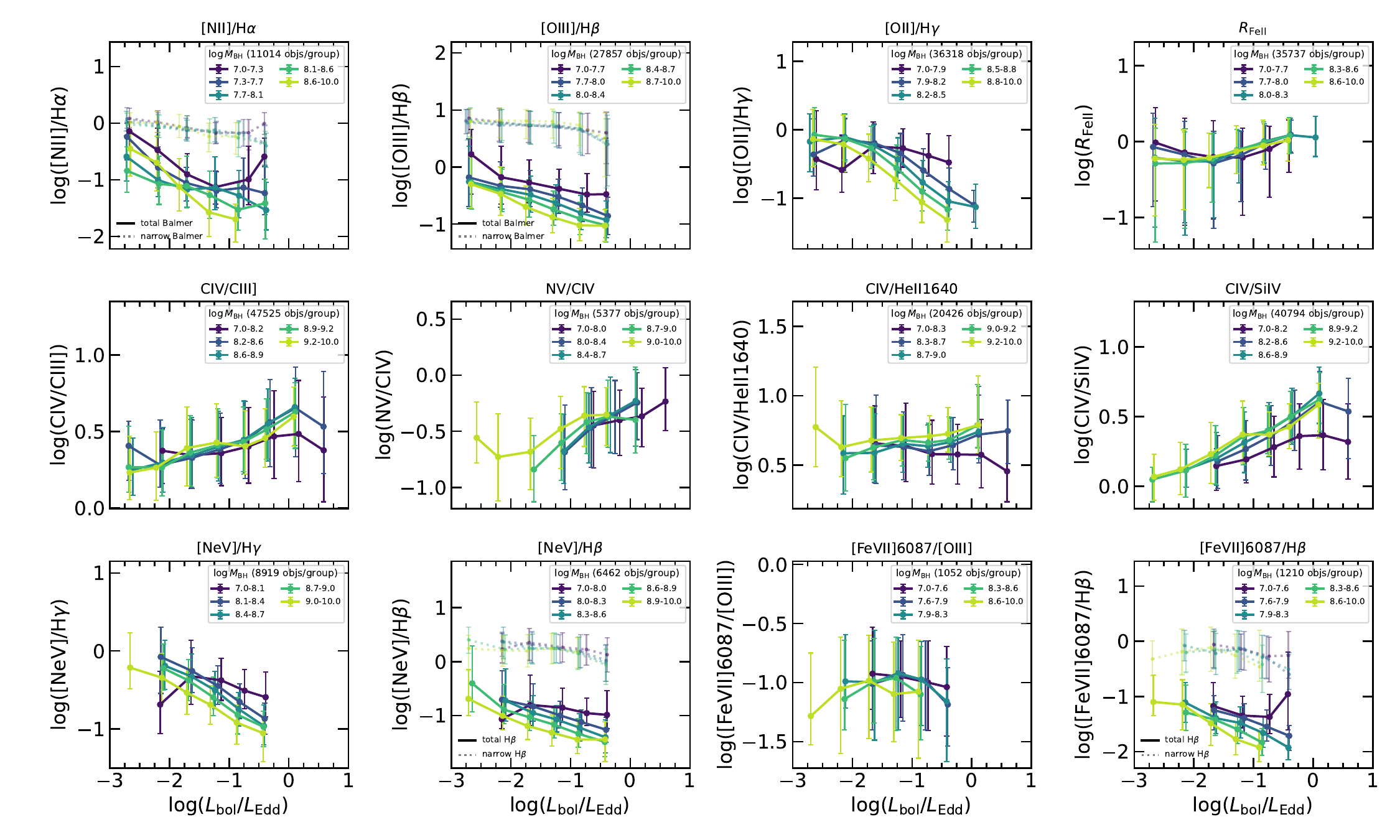}
    \caption{Emission-line ratios as functions of Eddington ratio (\LLEdd). In each panel, quasars are divided into quintiles of \Mbh\ with the mass intervals and the number of objects in each mass group given in the legends. The data points and error bars represent the median values and 16th--84th percentile range.}
    \label{fig:lineratio_LLEDD}
\end{figure*}

Emission-line ratios have long served as powerful probes of the ionizing SED in AGNs \citep{Kewley+2006}, most notably through the narrow-line BPT diagram \citep{BPT1981}. These narrow-line diagnostics become difficult to apply at high redshift, where aperture effects and the evolving ionization state of star-forming galaxies blur the classification boundaries that work well locally \citep{Kewley_etal2013, Steidel+2014, Arevalo+2025_JWST, Scholtz+2025}.
Broad emission lines (BELs) and CLs offer a different route, providing more direct constraints on the incident ionizing continuum than narrow lines alone, since the BLR reprocesses the central ionizing radiation and responds sensitively to continuum shape and intensity \citep{Baldwin_etal_1995, Korista&Goad_2000, McKaig_etal_2024}, and the long history of Eigenvector~1 studies demonstrates that the line strength between the optical \FeII\ and \hbeta\ already traces accretion rate reasonably well \citep{Boroson&Green1992, Sulentic_etal_2000, Shen&Ho2014}. 

Motivated by this, we test whether a broader set of optical and UV line ratios, spanning a range of ionization potentials, traces \LLEdd\ coherently across different line pairs. Figure~\ref{fig:lineratio_LLEDD} shows the emission-line ratios as a function of \LLEdd, with the sample divided into \Mbh\ quintiles to help disentangle accretion-rate trends from BH-mass dependence. The three rows present, respectively, the optical diagnostic ratios, UV BEL ratios, and CL-related ratios.
The top row contains the classic BPT narrow-line ratios \NII/\halpha\ and \OIII/\hbeta \citep{BPT1981, Veilleux&Osterbrock1987, Kewley+2006}, \OII/\hgamma, and $R_{\rm FeII}$, the Eigenvector~1 ratio of optical \FeII\ to broad \hbeta \citep{Boroson&Green1992, Sulentic_etal_2000, Shen&Ho2014}. Compared with the total (broad plus narrow) Balmer flux, \NII/\halpha, \OIII/\hbeta, and \OII/\hgamma\ all decline by roughly $\sim0.5$~dex toward higher \LLEdd, while $R_{\rm FeII}$ rises modestly, the same Eigenvector~1 behaviour in which \OIII weakens, and \FeII\ strengthens with increasing accretion rate \citep{Boroson&Green1992, Shen&Ho2014}. 
Replacing the total Balmer flux with the narrow component removes most of the decline: \NII/\halpha$_{\rm na}$ and \OIII/\hbeta$_{\rm na}$ remain near $-0.1$ and $0.55$, respectively, with only a weak decrease at the highest \LLEdd, in the same sense as the \LLEdd\ dependence of \NII/\halpha\ reported for X-ray selected AGN \citep{Oh+2017_BASS} and for low-redshift Type~1 AGN \citep{Stern&Laor2013}. The decline of the total-Balmer ratios is therefore driven mainly by the growing broad-line component, with only a modest change in the narrow-line ionization state. 
The middle row compares BELs that arise in the same region under comparable physical conditions, not split between the NLR and BLR. \CIV/\CIII and \CIV/\SiIV\ increase by $\sim0.3-0.5$~dex and \ion{N}{5}/\CIV\ rises by $\sim0.5$~dex with \LLEdd, whereas \CIV/\HeII\ is essentially flat. At higher accretion rates, \CIV\ is known to weaken in line strength and show stronger blueshifts \citep{Baskin&Laor2005, Richards+2011, Coatman+2016}. \citet{Temple+2021} show that quasars with large \CIV\ blueshifts have on average stronger \ion{N}{5}/\CIV\ and (\SiIV+\ion{O}{4}])/\CIV\ ratios, and once the \CIV\ blueshift is accounted for, these ratios show little dependence on Eddington ratio, suggesting that the line ratios are more closely tied to the outflow than to the accretion rate itself. These UV line ratios are also sensitive to the BLR photoionization conditions, including the ionization parameter, gas density, and incident SED, so their trends likely reflect a combination of outflow structure and changing BLR physical conditions.
The lack of a significant trend in \CIV/\HeII suggests that the hardness of the ionizing continuum near $50-55~\eV$ changes little over this range, so the trends are more likely governed by the gas conditions than by the SED shape alone.
In the bottom row with CLs, \NeV/\hgamma\ and \NeV/\hbeta\ decline with \LLEdd\ by $\sim0.6-0.7$~dex when compared to the total Balmer lines, and [\ion{Fe}{7}]~$\lambda$6087/\hbeta\ shows the same behaviour, matching the optical row, while the [\ion{Fe}{7}]/\OIII ratio is flat. The ratios between the CLs and the narrow Balmer lines (dotted curves) are also flat, indicating that the CL gas requiring photon energies of $\approx$100~eV or more (\NeV: 97~eV; [\ion{Fe}{7}]: 99~eV) tracks the NLR with no measurable change in the soft X-ray/EUV continuum hardness relative to the optical lines across our range. This is consistent with the lack of significant correlations between CL properties and Eddington ratio found in SDSS galaxies \citep{Reefe_etal_2022, Reefe_etal2023}, although that comparison is limited by detection incompleteness. 
Finally, the dependence on \Mbh\ is consistently weaker than the dependence on \LLEdd\ across these panels, and the sample is detection-limited at the lowest and highest accretion rates, and a significant CL detection requires strong lines in high-S/N spectra. 
Taken together, these ratios are sensitive to the ionizing SED in the EUV and soft X-ray regime, which cannot be observed directly for the vast majority of quasars. These additional line ratios may serve as a proxy for the accretion state of the central SMBH where a reliable single-epoch mass, and hence \LLEdd, is unavailable, including at high redshift, where only rest-frame UV lines fall within typical spectroscopic coverage. Reproducing the simultaneous trends across multiple, independent ratios with physically motivated AGN SEDs and photoionization modeling would place a considerably stronger constraint on the ionizing continuum shape than any single line ratio can provide in isolation \citep{Cann_etal_2018, WuQ_etal_2025}.

\subsection{Comparison with SDSS Quasar Catalogs}

\begin{figure*}
\centering
    \includegraphics[width=\linewidth]{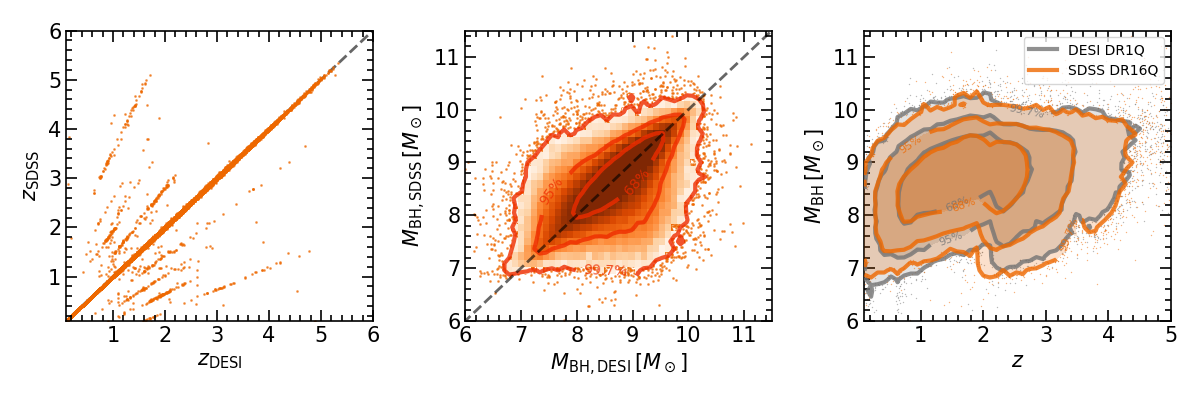}
    \caption{
    Comparison of redshift and virial BH mass measurements between DESI DR1 quasars and their SDSS DR16Q counterparts. 
    {Left}: Spectroscopic redshifts from DESI and SDSS;
    {Middle:} Virial BH masses derived from the two surveys. The contours correspond to the 1$\sigma$, 2$\sigma$, and 3$\sigma$ density levels of the distribution.
    {Right:} Distribution of black hole mass as a function of redshift for the matched quasars, with DESI measurements shown in gray and SDSS measurements in orange.}
    \label{fig:desi_vs_sdss}
\end{figure*}

\begin{figure*}
\centering
    \includegraphics[width=\linewidth]{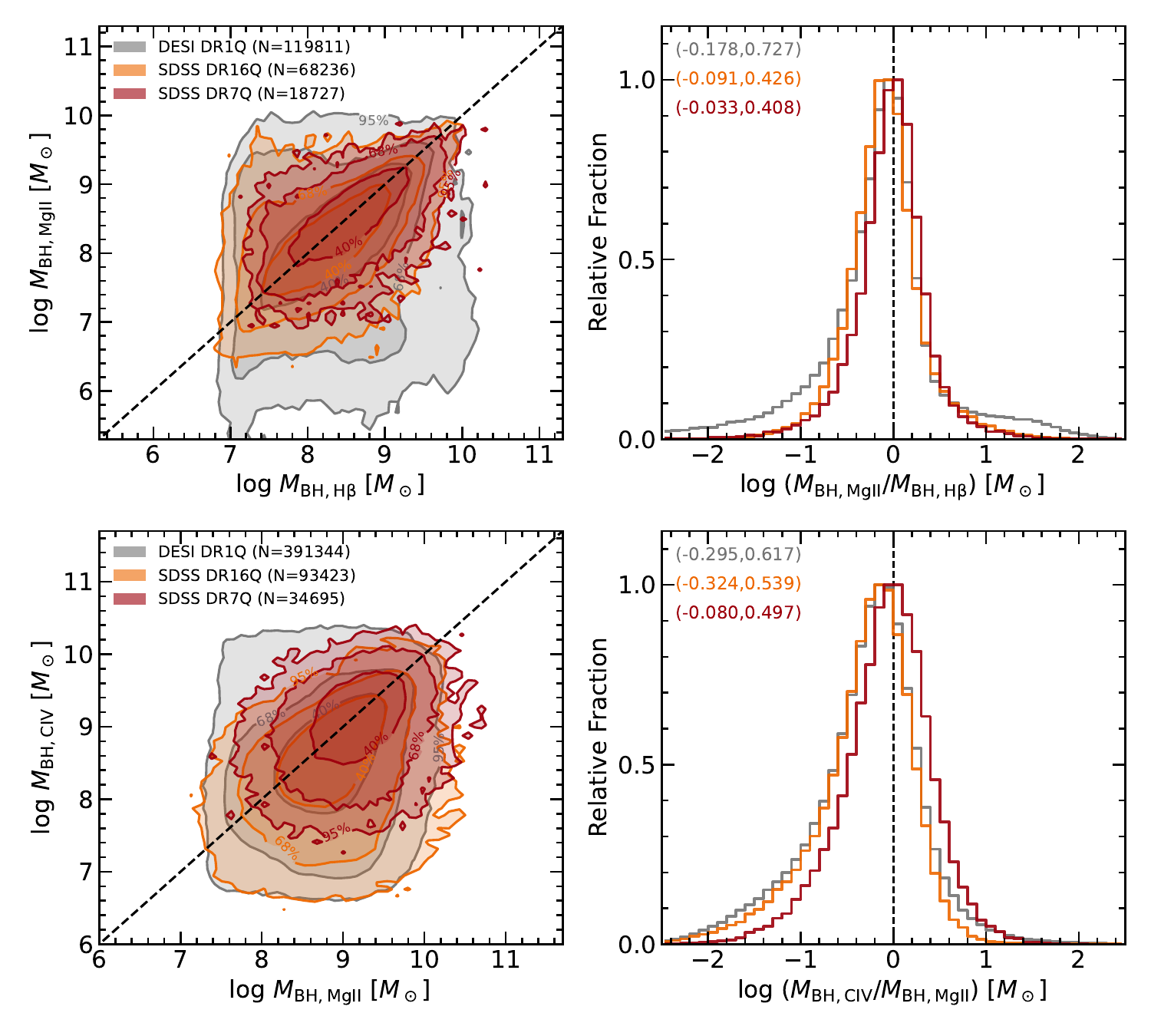}
    \caption{Comparison between single-epoch virial BH masses estimated from different BEL lines for quasars in DESI DR1 (black, constrained to {\tt SNR\_ALL\_MEDIAN}$>1$), SDSS DR16Q (orange), and SDSS DR7Q (brown). The mean ($\mu$) and standard deviation ($\sigma$) of the mass-difference distribution are shown in the top-left of the right panel. The contour levels correspond to the $40\%$, $68\%$, and $95\%$ density levels.}
    \label{fig:mbh_ratios}
\end{figure*}

We cross-match our DESI DR1 quasars to the SDSS DR16Q \citep{Wu&Shen2022} using a positional matching radius of $0.5$ arcsec. The matched sample contains 347,201 unique quasars, corresponding to $45.6\%$ of the SDSS DR16Q quasars and $20.2\%$ of the DESI DR1 catalog. Figure~\ref{fig:desi_vs_sdss} presents a comparison of the $z$ and \Mbh\ estimates derived from the two surveys.
The left panel shows the comparison of spectroscopic redshifts $z_{\rm sys}$ from both surveys. The vast majority of sources lie close to the one-to-one relation, demonstrating good overall consistency between the DESI and SDSS measurements. Only a small fraction of objects ($\lesssim0.2\%$) exhibit significant discrepancies with velocity offsets $|v_z|>5000\,\kms$. Most of these outliers occur in the redshift range $0.7\lesssim z\lesssim2.8$, where redshift determination relies primarily on broad rest-frame UV emission lines such as \CIV\ and \MgII. In this regime, misidentification of emission lines, template degeneracies in automated pipelines, or intrinsic velocity shifts of high-ionization lines can lead to systematic redshift differences. Our visual inspection of these sources suggests that DESI generally provides more reliable redshift estimates, potentially owing to higher spectral quality in these cases and the use of machine-learning-assisted quasar redshift validation \citep{Chaussidon_etal_2023_DESI, Alexander_etal_2023}.
In contrast, such discrepancies are uncommon at $z\lesssim1$, where narrow emission lines and host-galaxy stellar absorption features provide stronger constraints on the systemic redshift.
The middle and right panels compare the \Mbh\ measurements. The DESI masses broadly agree with those measured from SDSS spectra.

We test the internal consistency of our single-epoch BH mass estimates by comparing masses derived from different broad emission lines covered in the same spectrum (Figure~\ref{fig:mbh_ratios}). For reference, we show the corresponding distributions for SDSS DR16Q \citep{Wu&Shen2022} and SDSS DR7Q \citep{Shen_etal_2011}. In each catalog, we include all quasars with both relevant mass estimates rather than restricting the comparison to objects observed by multiple surveys. 
For the DESI sample, host-galaxy contamination is additionally mitigated using the spectral decomposition method of \citet{Ren_etal_2024}, which is particularly important for the optical continuum and \hbeta-based mass estimates at low redshift. Overall, the \MgII-based masses are consistent with those derived from \hbeta\ and \CIV, with the distributions centered close to the one-to-one relation. 
Some differences among the three catalogs are expected from the spectral-fitting methodology. In particular, SDSS DR7Q adopted local fits around individual broad emission lines. In contrast, DESI DR1 and SDSS DR16Q use global spectral fits, which can introduce small systematic differences in the measured BEL FWHM and fluxes. These offsets are nevertheless modest compared with the intrinsic systematic uncertainty of single-epoch virial BH mass estimators, which is typically of $\sim0.5$ dex, and additional uncertainty arises when recipes calibrated against local reverberation-mapped AGNs are extrapolated to luminous, distant quasars \citep{Shen2024_SDSSRM}.
The DESI DR1 distributions also show broader tails toward large discrepancies between the two mass estimates than those of DR16Q, most notably a tail of objects whose \CIV-based masses are much lower than their \MgII-based masses. 
Restricting the DESI sample to higher-S/N spectra substantially reduces these tails, as the objects in the tail are fainter and have lower S/N. Stacked spectra of these objects show a systematically narrower \CIV\ profile than those of the rest of the sample. We find that they lie at the low-luminosity end of the sample, and their \CIV\ profiles look noticeably narrower, possibly because of a prominent narrow \CIV\ component. The single-epoch mass recipe may therefore be unreliable in this regime.

The substantial overlap between the SDSS and DESI DR1 quasar samples provides a valuable opportunity to investigate quasar spectroscopic variability on a statistical basis. By presenting spectral measurements from both surveys within a consistent analysis framework, our data products enable systematic time-domain studies of quasars across a large, uniformly characterized sample.
Spectroscopic variability in quasars offers important insights into the structure and physics of the accretion flow and the BLR. For example, continuum variability can drive the “breathing” effect of the BLR, in which emission-line widths and strengths respond to changes in luminosity \citep{Korista&Goad_2004, SWang+2020}. Large spectroscopic surveys have also identified changing-look AGNs, where broad emission lines appear or disappear on timescales of years, likely reflecting significant changes in the accretion state or BLR structure \citep{Guo+2019_CLQ, Guo_etal_2020, Zeltyn+2024}. The multi-year baseline between SDSS and DESI is well suited for identifying such long-timescale spectral variability. Moreover, when combined with ongoing photometric surveys such as LSST and Euclid,  these datasets provide spectroscopic constraints that serve as a foundation for large statistical studies of quasar variability in the time domain.


\section{Conclusion} \label{sec:conclude}
In this work, we present a comprehensive spectroscopic analysis of the DESI DR1 quasar sample and construct a value-added catalog of quasar spectral properties based on uniform measurements of continuum and line emission. By applying a consistent fitting framework, we derive emission-line parameters, BH mass estimates, and improved systemic redshifts across a large and diverse quasar population. 
For 11.4\% of the spectra, our systemic redshifts deviate from the DESI pipeline redshifts by more than $500$~\kms and mitigate the biases associated with the blueshifts of high-ionization lines at $z>1$. Comparisons with SDSS quasar catalogs demonstrate overall consistency. We also provide local fits to ten coronal lines, with \NeV~$\lambda3426$ detected in 9.5\% of the spectra with coverage, and we show that some of the ratios of coronal, narrow, and broad lines vary systematically with the Eddington ratio.

The large sample size and homogeneous measurements provided by this catalog open new opportunities for statistical studies of quasar physics, including investigations of BLR structure, accretion properties, and emission-line diversity. Furthermore, the substantial overlap with previous spectroscopic surveys and synergy with ongoing and future time-domain and photometric programs make this dataset a valuable resource for studies of quasar variability and evolution. This VAC is designed to serve as a foundation for a wide range of applications, from precision redshift measurements to physically motivated modeling of AGN emission, and will be further expanded and refined in future DESI data releases.
All products from this work, including the catalog, input
fitting parameters, individual FITS outputs, Python notebooks, and update notes can be accessed from our data server \footnote{https://ariel.astro.illinois.edu/qsofit/DESI\_DR1/} and GitHub repository \footnote{https://github.com/QiaoyaWu/desi-dr1-quasar-vac}.

\begin{acknowledgments}

This work is partially supported by NSF grant AST-2509424.

This research used data obtained with the Dark Energy Spectroscopic Instrument (DESI). DESI construction and operations is managed by the Lawrence Berkeley National Laboratory. This material is based upon work supported by the U.S. Department of Energy, Office of Science, Office of High-Energy Physics, under Contract No. DE–AC02–05CH11231, and by the National Energy Research Scientific Computing Center, a DOE Office of Science User Facility under the same contract. Additional support for DESI was provided by the U.S. National Science Foundation (NSF), Division of Astronomical Sciences under Contract No. AST-0950945 to the NSF’s National Optical-Infrared Astronomy Research Laboratory; the Science and Technology Facilities Council of the United Kingdom; the Gordon and Betty Moore Foundation; the Heising-Simons Foundation; the French Alternative Energies and Atomic Energy Commission (CEA); the National Council of Humanities, Science and Technology of Mexico (CONAHCYT); the Ministry of Science and Innovation of Spain (MICINN), and by the DESI Member Institutions: www.desi.lbl.gov/collaborating-institutions. The DESI collaboration is honored to be permitted to conduct scientific research on I’oligam Du’ag (Kitt Peak), a mountain with particular significance to the Tohono O’odham Nation. Any opinions, findings, and conclusions or recommendations expressed in this material are those of the author(s) and do not necessarily reflect the views of the U.S. National Science Foundation, the U.S. Department of Energy, or any of the listed funding agencies.

This research uses services or data provided by the SPectra Analysis and Retrievable Catalog Lab (SPARCL) and the Astro Data Lab, which are both part of the Community Science and Data Center (CSDC) program at NSF National Optical-Infrared Astronomy Research Laboratory. NOIRLab is operated by the Association of Universities for Research in Astronomy (AURA), Inc. under a cooperative agreement with the National Science Foundation.

\end{acknowledgments}

\bibliography{refs}

\appendix
\begin{longtable*}{llll}
\caption[notes]{FITS Catalog Format}\label{tab:fits_catalog}\\
\hline \hline \\[0.2ex]
   \multicolumn{1}{l}{\textbf{Column}} &
   \multicolumn{1}{l}{\textbf{Format}} &
   \multicolumn{1}{l}{\textbf{Units}} &
   \multicolumn{1}{l}{\textbf{Description}} \\[0.2ex] \hline
   \\[0.2ex]
\endfirsthead
\multicolumn{4}{l}{{\tablename} \thetable{} -- Continued} \\[0.5ex]
  \hline \hline \\[0.2ex]
  \multicolumn{1}{l}{\textbf{Column}} &
  \multicolumn{1}{l}{\textbf{Format}} &
  \multicolumn{1}{l}{\textbf{Units}} &
  \multicolumn{1}{l}{\textbf{Description}} \\[0.2ex] \hline
  \\[0.2ex]
\endhead
  \hline
  \multicolumn{4}{l}{{Continued on Next Page\ldots}} \\
\endfoot
  \\[0.2ex] \hline \hline
\endlastfoot
 SURVEY & STRING &  & Survey name \\
 PROGRAM & STRING &  & DESI program type \\
 HEALPIX & LONG32 &  & Healpix number \\
 TARGETID & LONG64 &  & Unique identifier for each object observed by DESI \\
 Z\_AGNCAT & DOUBLE &  & Redshift provided by the DESI pipeline (AGN/QSO VAC) \\
 Z\_LYA & DOUBLE &  & Redshift from the \lya\ Y1 BAO analysis ($-1$ if not available) \\
 OBJID & STRING &  & Object identifier SURVEY-PROGRAM-HEALPIX-TARGETID \\
 RA & DOUBLE & degree & Right ascension (J2000) \\
 DEC & DOUBLE & degree & Declination (J2000) \\
 IN\_SDSS\_DR16Q & INT16 &  & 1 if cross-matched to SDSS DR16Q within 0.5\arcsec, 0 otherwise \\
 SDSS\_DR16Q\_OBJID & STRING &  & SDSS DR16Q object identifier (NA if no match) \\
 Z\_FIT & DOUBLE &  & Input redshift for PyQSOFit \\
 Z\_SYS & DOUBLE &  & Final systemic redshift (equals Z\_FIT where Z\_SYS\_ERR $=-1$) \\
 Z\_SYS\_ERR & DOUBLE &  & Uncertainty in Z\_SYS ($-1$ if Z\_FIT is adopted) \\
 Z\_SYS\_LINES & DOUBLE[8] &  & Systemic redshift from individual lines (broad \hbeta, \OIII, \CaII, \OII, \MgII, broad \CIII, \CIV, \SiIV) \\
 Z\_SYS\_LINES\_ERR & DOUBLE[8] &  & Uncertainties in Z\_SYS\_LINES \\
 SN\_RATIO\_CONTI & DOUBLE &  & Median S/N in the continuum windows \\
 SNR\_ALL\_MEDIAN & DOUBLE &  & Median S/N per pixel over the full spectrum \\
 EBV & DOUBLE & mag & Milky Way extinction $E(B-V)$ from \citet{SFD} \\
 CONTI\_PARA & DOUBLE[5] &  & Best-fit continuum parameters (PL normalization, slope, and polynomial coefficients) \\
 CONTI\_PARA\_ERR & DOUBLE[5] &  & Uncertainties in CONTI\_PARA \\
 FE\_UV\_OPT\_PARA & DOUBLE[6] &  & Best-fit UV and optical \FeII\ template parameters (normalization, FWHM, shift) \\
 FE\_UV\_OPT\_PARA\_ERR & DOUBLE[6] &  & Uncertainties in FE\_UV\_OPT\_PARA \\
 FE\_UV\_EW & DOUBLE[2] & \AA & Rest-frame EW of UV \FeII\ at 2250--2650 \AA, and its uncertainty \\
 FE\_OPT\_EW & DOUBLE[2] & \AA & Rest-frame EW of optical \FeII\ at 4434--4684 \AA, and its uncertainty \\
 LOGL1350 & DOUBLE & [erg s$^{-1}$] & Continuum luminosity $\log\lambda L_\lambda$ at rest-frame 1350 \AA \\
 LOGL1350\_ERR & DOUBLE & [erg s$^{-1}$] & Uncertainty in LOGL1350 \\
 LOGL1700 & DOUBLE & [erg s$^{-1}$] & Continuum luminosity $\log\lambda L_\lambda$ at rest-frame 1700 \AA \\
 LOGL1700\_ERR & DOUBLE & [erg s$^{-1}$] & Uncertainty in LOGL1700 \\
 LOGL2500 & DOUBLE & [erg s$^{-1}$] & Continuum luminosity $\log\lambda L_\lambda$ at rest-frame 2500 \AA \\
 LOGL2500\_ERR & DOUBLE & [erg s$^{-1}$] & Uncertainty in LOGL2500 \\
 LOGL3000 & DOUBLE & [erg s$^{-1}$] & Continuum luminosity $\log\lambda L_\lambda$ at rest-frame 3000 \AA \\
 LOGL3000\_ERR & DOUBLE & [erg s$^{-1}$] & Uncertainty in LOGL3000 \\
 LOGL4200 & DOUBLE & [erg s$^{-1}$] & Continuum luminosity $\log\lambda L_\lambda$ at rest-frame 4200 \AA \\
 LOGL4200\_ERR & DOUBLE & [erg s$^{-1}$] & Uncertainty in LOGL4200 \\
 LOGL5100 & DOUBLE & [erg s$^{-1}$] & Continuum luminosity $\log\lambda L_\lambda$ at rest-frame 5100 \AA \\
 LOGL5100\_ERR & DOUBLE & [erg s$^{-1}$] & Uncertainty in LOGL5100 \\
 CONTI\_FIT\_STAT & DOUBLE[2] &  & Number of pixels and reduced $\chi^2$ of the continuum fit \\
 FRAC\_HOST\_4200 & DOUBLE &  & Host-galaxy fraction at rest-frame 4200 \AA \\
 FRAC\_HOST\_5100 & DOUBLE &  & Host-galaxy fraction at rest-frame 5100 \AA \\
 DN4000 & DOUBLE &  & $D_{\rm n}4000$ index of the host galaxy \\
 HOST\_DECOMP\_PARA & DOUBLE[5] &  & Coefficients of the five eigenspectra used in the host decomposition \\
\hline
   &   &  \AA, $10^{-17}$ erg s$^{-1}$ cm$^{-2}$, & peak wavelength, flux, \\
   &   & [erg s$^{-1}$], km s$^{-1}$, \AA, \AA &  $\log L_{\rm line}$, FWHM, EW, 50\% flux centroid wavelength \\
 HALPHA & DOUBLE[6] & \nodata & For the entire \halpha\ profile (narrow and broad combined) \\
 HALPHA\_BR & DOUBLE[6] & \nodata & For the broad \halpha\ component \\
 HALPHA\_NA & DOUBLE[6] & \nodata & For the narrow \halpha\ component \\
 NII6549 & DOUBLE[6] & \nodata & For the narrow [N\,{\sc ii}]\,$\lambda$6549 component \\
 NII6585 & DOUBLE[6] & \nodata & \nodata \\
 SII6718 & DOUBLE[6] & \nodata & \nodata \\
 SII6732 & DOUBLE[6] & \nodata & \nodata \\
 HBETA & DOUBLE[6] & \nodata & For the entire \hbeta\ profile (narrow and broad combined) \\
 HBETA\_BR & DOUBLE[6] & \nodata & For the broad \hbeta\ component \\
 HBETA\_NA & DOUBLE[6] & \nodata & For the narrow \hbeta\ component ($z<1.02$ only) \\
 HEII4687 & DOUBLE[6] & \nodata & For the entire He\,{\sc ii}\,$\lambda$4687 profile (narrow and broad combined) \\
 HEII4687\_BR & DOUBLE[6] & \nodata & For the broad He\,{\sc ii}\,$\lambda$4687 component \\
 OIII5007 & DOUBLE[6] & \nodata & For the entire [O\,{\sc iii}]\,$\lambda$5007 profile \\
 OIII5007C & DOUBLE[6] & \nodata & For the core [O\,{\sc iii}]\,$\lambda$5007 profile \\
 OIII4960 & DOUBLE[6] & \nodata & For the entire [O\,{\sc iii}]\,$\lambda$4960 profile \\
 OIII4960C & DOUBLE[6] & \nodata & For the core [O\,{\sc iii}]\,$\lambda$4960 profile \\
 HGAMMA & DOUBLE[6] & \nodata & \nodata \\
 HDELTA & DOUBLE[6] & \nodata & \nodata \\
 MGII & DOUBLE[6] & \nodata & For the entire \MgII\ profile (narrow and broad combined) \\
 MGII\_BR & DOUBLE[6] & \nodata & For the broad \MgII\ component \\
 CIII\_BR & DOUBLE[6] & \nodata & For the broad \CIII\ component \\
 SIIII1892 & DOUBLE[6] & \nodata & \nodata \\
 ALIII1857 & DOUBLE[6] & \nodata & \nodata \\
 NIII1750 & DOUBLE[6] & \nodata & \nodata \\
 CIV & DOUBLE[6] & \nodata & \nodata \\
 HEII1640 & DOUBLE[6] & \nodata & For the entire He\,{\sc ii}\,$\lambda$1640 profile (narrow and broad combined) \\
 HEII1640\_BR & DOUBLE[6] & \nodata & For the broad He\,{\sc ii}\,$\lambda$1640 component \\
 SIIV\_OIV & DOUBLE[6] & \nodata & For the 1400 \AA\ complex \\
 OI1304 & DOUBLE[6] & \nodata & \nodata \\
 LYA & DOUBLE[6] & \nodata & \nodata \\
 NV1240 & DOUBLE[6] & \nodata & \nodata \\
 NEV3346 & DOUBLE[6] & \nodata & Local fit (linear continuum + one Gaussian) of [Ne\,{\sc v}]\,$\lambda$3346 \\
 NEV3426 & DOUBLE[6] & \nodata & Local fit of [Ne\,{\sc v}]\,$\lambda$3426 \\
 FEVII3586 & DOUBLE[6] & \nodata & Local fit of [Fe\,{\sc vii}]\,$\lambda$3586 \\
 OII3728 & DOUBLE[6] & \nodata & Local fit of [O\,{\sc ii}]\,$\lambda\lambda$3727,3729 \\
 FEVII3759 & DOUBLE[6] & \nodata & Local fit of [Fe\,{\sc vii}]\,$\lambda$3759 \\
 CAII3934 & DOUBLE[6] & \nodata & For the \CaII\ K absorption line (local fit; negative flux and EW) \\
 FEXIV5303 & DOUBLE[6] & \nodata & Local fit of [Fe\,{\sc xiv}]\,$\lambda$5303 \\
 FEVII5721 & DOUBLE[6] & \nodata & Local fit of [Fe\,{\sc vii}]\,$\lambda$5721 \\
 FEVII6087 & DOUBLE[6] & \nodata & Local fit of [Fe\,{\sc vii}]\,$\lambda$6087 \\
 FEX6374 & DOUBLE[6] & \nodata & Local fit of [Fe\,{\sc x}]\,$\lambda$6374 \\
 SXII7611 & DOUBLE[6] & \nodata & Local fit of [S\,{\sc xii}]\,$\lambda$7611 \\
 FEXI7892 & DOUBLE[6] & \nodata & Local fit of [Fe\,{\sc xi}]\,$\lambda$7892 \\
\hline
 &   &  \AA, $10^{-17}$ erg s$^{-1}$ cm$^{-2}$, & Uncertainties in peak wavelength, flux, \\
 &   & [erg s$^{-1}$], km s$^{-1}$, \AA, \AA &  $\log L_{\rm line}$, FWHM, EW, 50\% flux centroid wavelength \\
 HALPHA\_ERR & DOUBLE[6] & \nodata & \nodata \\
 HALPHA\_BR\_ERR & DOUBLE[6] & \nodata & \nodata \\
 HALPHA\_NA\_ERR & DOUBLE[6] & \nodata & \nodata \\
 NII6549\_ERR & DOUBLE[6] & \nodata & \nodata \\
 NII6585\_ERR & DOUBLE[6] & \nodata & \nodata \\
 SII6718\_ERR & DOUBLE[6] & \nodata & \nodata \\
 SII6732\_ERR & DOUBLE[6] & \nodata & \nodata \\
 HBETA\_ERR & DOUBLE[6] & \nodata & \nodata \\
 HBETA\_BR\_ERR & DOUBLE[6] & \nodata & \nodata \\
 HBETA\_NA\_ERR & DOUBLE[6] & \nodata & \nodata \\
 HEII4687\_ERR & DOUBLE[6] & \nodata & \nodata \\
 HEII4687\_BR\_ERR & DOUBLE[6] & \nodata & \nodata \\
 OIII5007\_ERR & DOUBLE[6] & \nodata & \nodata \\
 OIII5007C\_ERR & DOUBLE[6] & \nodata & \nodata \\
 OIII4960\_ERR & DOUBLE[6] & \nodata & \nodata \\
 OIII4960C\_ERR & DOUBLE[6] & \nodata & \nodata \\
 HGAMMA\_ERR & DOUBLE[6] & \nodata & \nodata \\
 HDELTA\_ERR & DOUBLE[6] & \nodata & \nodata \\
 MGII\_ERR & DOUBLE[6] & \nodata & \nodata \\
 MGII\_BR\_ERR & DOUBLE[6] & \nodata & \nodata \\
 CIII\_BR\_ERR & DOUBLE[6] & \nodata & \nodata \\
 SIIII1892\_ERR & DOUBLE[6] & \nodata & \nodata \\
 ALIII1857\_ERR & DOUBLE[6] & \nodata & \nodata \\
 NIII1750\_ERR & DOUBLE[6] & \nodata & \nodata \\
 CIV\_ERR & DOUBLE[6] & \nodata & \nodata \\
 HEII1640\_ERR & DOUBLE[6] & \nodata & \nodata \\
 HEII1640\_BR\_ERR & DOUBLE[6] & \nodata & \nodata \\
 SIIV\_OIV\_ERR & DOUBLE[6] & \nodata & \nodata \\
 OI1304\_ERR & DOUBLE[6] & \nodata & \nodata \\
 LYA\_ERR & DOUBLE[6] & \nodata & \nodata \\
 NV1240\_ERR & DOUBLE[6] & \nodata & \nodata \\
 NEV3346\_ERR & DOUBLE[6] & \nodata & \nodata \\
 NEV3426\_ERR & DOUBLE[6] & \nodata & \nodata \\
 FEVII3586\_ERR & DOUBLE[6] & \nodata & \nodata \\
 OII3728\_ERR & DOUBLE[6] & \nodata & \nodata \\
 FEVII3759\_ERR & DOUBLE[6] & \nodata & \nodata \\
 CAII3934\_ERR & DOUBLE[6] & \nodata & \nodata \\
 FEXIV5303\_ERR & DOUBLE[6] & \nodata & \nodata \\
 FEVII5721\_ERR & DOUBLE[6] & \nodata & \nodata \\
 FEVII6087\_ERR & DOUBLE[6] & \nodata & \nodata \\
 FEX6374\_ERR & DOUBLE[6] & \nodata & \nodata \\
 SXII7611\_ERR & DOUBLE[6] & \nodata & \nodata \\
 FEXI7892\_ERR & DOUBLE[6] & \nodata & \nodata \\
\hline
 Ha\_STAT & DOUBLE[2] &  & Number of pixels, reduced $\chi^2$ of the line-complex fit \\
 Hb\_STAT & DOUBLE[2] &  & \nodata \\
 HgHd\_STAT & DOUBLE[2] &  & \nodata \\
 MgII\_STAT & DOUBLE[2] &  & \nodata \\
 CIII\_STAT & DOUBLE[2] &  & \nodata \\
 CIV\_STAT & DOUBLE[2] &  & \nodata \\
 SiIV\_STAT & DOUBLE[2] &  & \nodata \\
 Lya\_STAT & DOUBLE[2] &  & \nodata \\
 NEV3346\_STAT & DOUBLE[2] &  & \nodata \\
 NeV\_STAT & DOUBLE[2] &  & \nodata \\
 FEVII3586\_STAT & DOUBLE[2] &  & \nodata \\
 OII\_STAT & DOUBLE[2] &  & \nodata \\
 FEVII3759\_STAT & DOUBLE[2] &  & \nodata \\
 CaII\_STAT & DOUBLE[2] &  & \nodata \\
 FEXIV5303\_STAT & DOUBLE[2] &  & \nodata \\
 FEVII5721\_STAT & DOUBLE[2] &  & \nodata \\
 FEVII6087\_STAT & DOUBLE[2] &  & \nodata \\
 FEX6374\_STAT & DOUBLE[2] &  & \nodata \\
 SXII7611\_STAT & DOUBLE[2] &  & \nodata \\
 FEXI7892\_STAT & DOUBLE[2] &  & \nodata \\
\hline
 LOGLBOL & DOUBLE & [erg s$^{-1}$] & Bolometric luminosity \\
 LOGLBOL\_ERR & DOUBLE & [erg s$^{-1}$] & Uncertainty in LOGLBOL \\
 LOGMBH\_HB & DOUBLE & [M$_\odot$] & Single-epoch BH mass based on \hbeta \\
 LOGMBH\_HB\_ERR & DOUBLE & \nodata & Uncertainty in LOGMBH\_HB \\
 LOGMBH\_MGII & DOUBLE & \nodata & Single-epoch BH mass based on \MgII \\
 LOGMBH\_MGII\_ERR & DOUBLE & \nodata & \nodata \\
 LOGMBH\_CIV & DOUBLE & \nodata & Single-epoch BH mass based on \CIV \\
 LOGMBH\_CIV\_ERR & DOUBLE & \nodata & \nodata \\
 LOGMBH & DOUBLE & \nodata & Fiducial single-epoch BH mass \\
 LOGMBH\_ERR & DOUBLE & \nodata & \nodata \\
 LOGLLEDD\_RATIO & DOUBLE &  & Eddington ratio based on the fiducial BH mass \\
 LOGLLEDD\_RATIO\_ERR & DOUBLE &  & Uncertainty in LOGLLEDD\_RATIO \\
 BAL\_FLAG & INT16 &  & BAL flag from \ion{C}{4} BI: 1 = BAL (BI$>0$), 0 = non-BAL, $-1$ = not in the DESI QSO catalog (main/dark, $z>1.57$ only) \\
\end{longtable*}

\end{document}